# Thin Dielectric Films in Superconducting Cavities

## E. L. Garwin and Mario Rabinowitz

*Stanford Linear Accelerator Center, Stanford University, Stanford, California 94305*

Inquiries to: *Armor Research, 715 Lakemead Way, Redwood City, CA 94062*
Mario715@earthlink.net

### Abstract

At high field levels, field emission losses in superconducting cavities have an adverse effect in both reducing the otherwise extremely high $Q > 10^9$ , and the magnetic breakdown field bringing it well below the critical magnetic field of the super-conductor. The two effects may well be related as field enhancing whiskers responsible for the field emission losses can be driven normal thus reducing the magnetic breakdown field. Both the beneficial and deleterious effects of anodizing the inner surface of superconducting cavities are discussed and analyzed.

It has been reported [1, 2] that anodic oxidation of Nb cavities leads to an improvement by a factor of ~ 2 in both the superconducting $Q$ and Magnetic breakdown field $H_p$'. Coating the surface of superconducting cavities with a dielectric was proposed both to reduce field emission losses and to protect the cavity froin deleterious effects of air exposure [3]. Experiments carried out [3] to determine the effect on field emission of dielectric films showed that Pb whiskers could grow through an $Al_2O_3$ overcoating and vitiate any reduction of the electric field by the dielectric.

We propose that rather than merely adding to the metal a dielectric layer which reduces the local electric field by a factor k (the dielectric constant), anodization of niobium converts small conducting protrusions into dielectric material, thus vastly reducing or entirely eliminating their ability to enhance both electric and magnetic fields. This mechanism is especially important for the Martens *et al.* [1] cavity which was operated in the $TE_{011}$ mode, with no electric field on the cavity surface. Our suggestion [4] has generated interest in the possibility that successive anodization and stripping of the oxide may be a partial substitute for the technique of high-temperature firing of cavities to improve their Q and $H_p$'. This may allow the use of cheaper, lower-temperature processing ovens. We feel that the presence of an anodic dielectric such as $Nb_2O_5$, may ultimately lead to deleterious effects in a superconducting cavity.

Since the reported work (1,2) made no mention of the loss tangent tg $\delta$ of $Nb_2O_5$ and it is not available in the published literature, we here make an upper-limit calculation. The average dielectric power loss per unit volume is

$$\left\langle \frac{dP}{dV} \right\rangle \sim \frac{1}{2}\,\omega k \varepsilon_o E_o^2 \mathrm{tg}\,\delta \tag{1}$$

where $\omega$ is the angular frequency, k is the dielectric constant, $\varepsilon_o$ is the vacuum permitivity, and $E_o$ is the peak clectric field at the dielectric. Although the addilion of $Nb_2O_5$ increased Q slightly, to put an upper limit on tg $\delta$ we will assume that the dielectric loss dominates in the oxide-coated cavity

$$Q \approx \frac{2\pi\left[\frac{1}{4}\int_0^{V_c}\left(\mu_o H_o^2 + \varepsilon_o E_o^2\right)dV\right]}{\frac{2\pi}{\omega}\int_0^{V_d}\frac{1}{2}\omega k \varepsilon_o E_o^2 \mathrm{tg}\,\delta\, dV}, \tag{2}$$

where $V_c$ and $V_d$ are, respectively, the volume of the cavity and of the dielectric, $\mu_o$ is the vacuum permeability, and $H_o$ is the peak magnetic field..

Therefore, to a good approximation for TM cavities,

$$k\,tg\,\delta \sim \frac{1}{Q}\left(\frac{V_c}{V_d}\right). \tag{3}$$

Kneisel et al. [2] report $Q \sim 10^9$ (at an unspecified temperature, probably $\sim$ 2 K) for their *TE$_{011}$* S-band cavity with $\left(\dfrac{V_c}{V_d}\right) \sim 10^5$. Therefore, from eq. (3),

$k\,tg\,\delta \sim 10^{-4}$ . Extrapolating the data of Duffy et al. [5], $k \approx 6$ at 2.6 GHz assuming no resonance in this region. Therefore, for $Nb_2O_5$, $tg\,\delta \sim 10^{-4}$ . Extrapolating the data of Duffy et al. [5], $k \approx 6$ at 2.6 GHz assuming no resonance in this region. Therefore, for $Nb_2O_5$, $tg\,\delta \sim 10^{-5}$ at 2.6 GHz and $T \sim 2$ K.

It should be borne in mind that $Nb_2O_5$ is an ionic dielectric [6]. Because of the distribution of relaxation times in these materials, there is no sound basis for extrapolation of the value of the loss tangent calculated above to considerably different frequencies and temperatures. For example, the loss tangent of another ionic dielectric, NaCl, has been measilre [7] over, the frequency range 40 to 1000 MHz. The loss tangent shows rapid variations with frequency and temperature. The 4.2K $tg\,\delta$ values increase from $3.2 \times 10^{-5}$ at 434 MHz to $9.7 \times 10^{-5}$ at 996 MHz. The loss tangent at, 996 MHz ($9.7 \times 10^{-5}$) decereases to $5.3 \times 10^{-5}$ at 2.2K and further to $3.2 \times 10^{-5}$ at 1.8K. Thesc dramatic and rapid variations indicate that extreme caution should be used in extrapolating results obtained for $Nb_2O_5$ at 2.6 GHz and $\sim$ 2K to other frequencies and temperatures, partieularly L-band ($\sim$1.3 GHz) at which a large inachine is presently being constructed.

It has been established that irradiation of diclectrics [8] can inerease, the low-temperature value, of the loss tangent by a factor between 10 and 100. Oxide-coated cavities in an acclerator would certainly find themselves in a radiation environment.  A reduction of Q by a factor of 10 during operation would probably not be acceptable . The dielectric with the lowest loss prior to irradiation may not be the most radiation resistant.

It is well known that, color centers may be formed in dieltectrics irradiated by photons of enerlgy a few tens of eV and higher, as well as by particle bombardment.  A color center is a lattice, defect (binding a charged particle), which absorbs light. Therefore (aside from radiation produced by beam missteering and accidents), field-emitted electrons as well as the synchrotron radiation concomitant with necessary bending of the electrion beam (as in a recirculator) will produce charge traps in many dielectrics.  The interaction of the bound charge in these defects with the electric field will greatly increase the dielectric loss as has been observed for quartz [8].

To minimize the effects of a large dielectric loss, one might be tempted to make the dielectric thinner. Aside from the problenis of making a continuous dielectric film which is substantially thinner (30 Å rather than 300 Å), one encounters the Malter [9], Stern, Gossling and Fowler [10] effect. If the dielectric charges positively (due to a secondary-electron emission coetficient greater than one, and/or positive ion impact), then, for thin films, the electric field across the layer can produce copious electron ernission from the substrate by a tunneling process [11]. This is an additional source of power loss. For thick films, the electric field usually does not build up to a higlh enough value for this to happen, but dielectric voltage breakdown or substantial conduction losses may occur.

The presence of the dielectric enhances the secondary- electron emission yield over that of the metal, further increasing the likelihood of multipactoring and associated problems.

In addition to reducing field emission, a dielectric coating on a *cathode* may also increase the breakdown voltage [3]. Jedynak [12] has increased the breakdown voltage by almost a factor of 2 by using dielectric films thicker than 2000 Å on his *cathodes.* However, a dielectric coating on an *anode* can severely lower the breakdown voltage. For radiofreqnency fields, the net effect on breakdown voltage due to dielectric coatings is yet to be determined. If the coating is not self-healing, a breakdown which punctures the dielectric will generally cause subsequent breakdown at voltages lower than if the dielectric were not present at all [12].

Stress induced, during thermal cycling, due to lattice mismatch between dielectric and substrate can enhance the probability of whisker growth. As was observed [3] whiskers were capable of growing right through the dielectric.

Though the addition of a dielectric coating might possibly be advantageous, unless the above effects are understood and studied, any overall advantageous result may turn out to be serendipitous.

**References**

1. H. Martens, H. Diepers and R. K. Sun: *Phys. Lett.,* **34A**, 439 (1971).

2. P. Kneinsel, 0. Stoltz And J. Halbritter: *1971 National Particle Accelerator Conference, Washington, D. C.*

3. *M.* Rabinowrrz and E. L. Garwin: *High-clectric-field effects in a superconducting accelerator,* Report No. SLAC-TN-68-27 (1968).

4 E. L. Garwin and M. Rabinowitz: *Thin-film dielectric power losses in superconducting cavities,* Report No. SLAC-TN-71-9 (1971).

5.  M. T. Duffy, C.C. Wang, A. Waxman, and K. H. Zaininger: *Journ.Electrochcm. Soc.: Solid State Science*, **116,** 234 (1969).

6.  M. L. A. Robinson and H Roetschi: J*ourn. Phys. Chem. Solids*, **29**, 1503 (1968).

7.  D. Grissom and W. H. Hartwig: Theory and measurement of dielectric propolies of alkalhi haliide crystals at cryogenic temperatures, University ot Texas, Austin Lab. Tech. Report 6 (1965).

8.  J.  Volger, J. M. Stevens, C. Von Amerongen: *Philips Res. Rcp.*, **10**, 260 (1955).

9.  L. Malter: *Phys. Rev.*, 50, 48 (1936).

10.   T. E. Stern,  B. S. Gossmng and R. H. Fowler: *Proc. Rov. Soc.*, **124A**, 699 (1929).

11.  M. Rabinowitz: Electrical Conductivity in. High Vacuum, Report No. 81,AC-TN-68-23 (1968);  Analysis of Critical Power Loss in a Superconductor. *J. Appl. Phys. **42**, 88-96 (1971).*

12.  L. Jedynak: Journ. Appl. Phys. **35**, 1727 (1964).